\documentclass[aps,prb,twocolumn,showpacs,groupedaddress]{revtex4}
\usepackage{graphicx}

\begin{document}

\title{Pressure-induced polarization reversal in multiferroic $YMn_2O_5$}
\author{Rajit P. Chaudhury$^{1}$, Clarina R. dela Cruz$^{2,3}$, Bernd Lorenz$^{1}$, Yanyi Sun$^{1}$, Ching-Wu Chu$^{1,4,5}$, S. Park$^{6}$, and Sang-W. Cheong$^{6}$}
\affiliation{$^{1}$Department of Physics and TCSUH, University of Houston, Houston, TX 77204, USA} \affiliation{$^{2}$Department of Physics and
Astronomy, University of Tennessee, Knoxville, TN 37996, USA} \affiliation{$^3$Oak Ridge National Laboratory, Oak Ridge, Tennessee 37831, USA}
\affiliation{$^{4}$Lawrence Berkeley National Laboratory, 1 Cyclotron Road, Berkeley, CA 94720, USA} \affiliation{$^{5}$Hong Kong University of
Science and Technology, Hong Kong, China} \affiliation{$^{6}$Rutgers Center for Emergent Materials and Department of Physics and Astronomy,
Rutgers University, Piscataway, NJ 08854, USA}
\date{\today }

\begin{abstract}
The low-temperature ferroelectric polarization of multiferroic $YMn_2O_5$ is completely reversed at a critical pressure of 10 kbar and the phase
transition from the incommensurate to the commensurate magnetic phase is induced by pressures above 14 kbar. The high-pressure data correlate
with thermal expansion measurements indicating a significant lattice strain at the low-temperature transition into the incommensurate phase. The
results support the exchange striction model for the ferroelectricity in multiferroic $RMn_2O_5$ compounds and they show the importance of
magnetic frustration as well as the spin-lattice coupling.
\end{abstract}

 \maketitle











Multiferroic magnetoelectric compounds in which ferroelectricity and magnetic order coexist and mutually interact have been in the focus of
attention very recently because of a wealth of novel physical phenomena observed in these complex
materials.\cite{fiebig:05,eerenstein:06,cheong:07,rao:07} Among the family of multiferroics the $RMn_2O_5$ ($R$ = rare earth, Y, Bi) manganites
are of significant interest because of their extreme complexity with transitions between different commensurate (CM) and incommensurate (IC)
magnetic structures some of which exhibit ferroelectricity induced by frustrated magnetic
orders.\cite{inomata:96,hur:04,blake:05,kobayashi:04,ratcliff:05,chapon:06} Their common phase sequence upon decreasing temperature, $T$,
includes transitions from the high-$T$ paramagnetic and paraelectric (PE) to an antiferromagnetic (AFM) but still PE phase at $T_{N1}\simeq$ 44
K with an IC magnetic modulation along the orthorhombic a- and c-axes. At $T_{C1}\simeq$ 39 K, a lock-in transition takes place into a CM
magnetic phase ($\overrightarrow{q}=(0.5,0,0.25)$) which is ferroelectric (FE). At lower temperature, $T_{C2}$, all rare earth $RMn_2O_5$
compounds experience another transition where $\overrightarrow{q}$ unlocks again into different IC values and the FE polarization shows a sharp
drop (but not necessarily to zero). Additional changes of the magnetic order identified as spin re-orientations in the CM phase have been
reported for some $RMn_2O_5$, for example $DyMn_2O_5$\cite{ratcliff:05} and $HoMn_2O_5$.\cite{delacruz:06} It is remarkable that all magnetic
(and FE) phase transitions are clearly resolved in anomalies of the dielectric constant, the FE polarization, and the thermal expansivities
which proves the prominent role of the spin-lattice coupling in the multiferroic properties of these compounds.\cite{delacruz:06,delacruz:06c}
The origin of the phase complexity of $RMn_2O_5$ lies in the peculiarities of the lattice structure with different magnetic ions ($Mn^{4+}$,
$Mn^{3+}$, rare earth) and a multitude of partially competing superexchange interactions leading to a high degree of frustration in the magnetic
system.\cite{blake:05} In particular, the smallest loop of nearest-neighbor Mn ions with AFM exchange interactions in the $a-b$ plane has five
members which naturally leads to geometric frustration of the Mn spins.\cite{delacruz:06} Frustrated systems are susceptible to small
perturbations like magnetic fields or pressure since many different magnetic structures are close in energy and compete for the ground state.
For some $RMn_2O_5$ compounds it was recently demonstrated that the low-temperature IC phase is extremely sensitive to magnetic
fields\cite{higashiyama:04,ratcliff:05,higashiyama:05,delacruz:06c} or to hydrostatic pressure.\cite{delacruz:07} This opens not only new venues
to mutually control magnetic and FE properties but it also provides important insight into the microscopic interactions that are essential for
the physics of multiferroic materials.

$YMn_2O_5$ is unique among all $RMn_2O5$ compounds in that it experiences a FE polarization reversal at $T_{C2}$, the transition into the low-T
incommensurate phase.\cite{inomata:96,chapon:06} This effect was attributed to a change of the relative phase of the AFM orders of neighboring
(a-axis) chains of $Mn^{3+}-Mn^{3+}-Mn^{4+}-Mn^{3+}-Mn^{3+}-Mn^{4+}$... spins based on an exchange striction mediated mechanism for the FE
displacement of frustrated $Mn^{3+}-Mn^{4+}$ spins between neighboring chains.\cite{chapon:06,cheong:07,delacruz:06} However, recent experiments
have detected a small deviation of the Mn-spins from the $a-b$ plane forming a non-collinear (spiral) modulation along the $c$-axis in
$RMn_2O_5$ multiferroics.\cite{noda:06,kimura:06,kimura:07,vecchini:07} This spiral spin order also breaks the inversion
symmetry\cite{mostovoy:06,cheong:07} and it was considered as an alternative mechanism for ferroelectricity in these
compounds.\cite{noda:06,kimura:06,kimura:07,okamoto:07} The essential difference between the two proposed mechanisms is the microscopic nature
of the exchange coupling that has to be involved. In the spin current (spiral) model\cite{katsura:05,sergienko:06} the weak antisymmetric
Dzyaloshinskii-Moriya interaction ($\sim\overrightarrow{S_i}\times\overrightarrow{S_j}$) creates the magnetoelectric coupling whereas in the
exchange striction model the FE polarization has its origin in the ionic displacements lifting the frustration of spins interacting via the
symmetric exchange (or superexchange) interactions ($\sim\overrightarrow{S_i}\cdot\overrightarrow{S_j}$). Whether the spiral spin model or the
exchange striction model provides the better explanation of the FE order in the $RMn_2O_5$ compounds is still a matter of
discussion\cite{kim:08,radaelli:08} and needs further explorations. In order to gain a deeper insight into the microscopic mechanisms of
ferroelectricity of $RMn_2O_5$ we have investigated the effect of hydrostatic pressure on the FE properties of $YMn_2O_5$ and found that the FE
polarization in the low-temperature IC phase ($T<T_{C2}$) reverses sign above a critical pressure. These data correlate with the lattice strain
at $T_{C2}$ measured by high resolution thermal expansion experiments. Our results are interpreted within the exchange striction model in terms
of a pressure control of the relative phase of magnetic orders in neighboring AFM zigzag chains.

Single crystals of $YMn_2O_5$ grown from the flux were polished to thin platelets (0.2 mm thick) so that the dielectric constant and the
pyroelectric current could be measured along the orthorhombic $b$-axis. Hydrostatic pressure was applied employing a beryllium-copper clamp
cell.\cite{delacruz:07} A mixture of Fluorinert 70 and 77 liquids was used as the pressure transmitting medium. Thermal expansion measurements
were conducted along the $a$-, $b$-, and $c$-axes using a capacitance dilatometer. Since $YMn_2O_5$ exhibits a polarization reversal at $T_{C2}$
special care was taken to measure the FE polarization correctly in both FE phases, above and below $T_{C2}$, respectively. Details of the
experimental procedure are discussed elsewhere.\cite{delacruz:08}

The dielectric constant and FE polarization at ambient pressure, shown in Fig. 1, is consistent with earlier
reports,\cite{inomata:96,matsumoto:03,kagomiya:05} however, a careful investigation reveals subtle features that have to be discussed in more
detail. Upon decreasing temperature the dielectric constant shows a sudden increase at $T_{N1}$, the onset of sinusoidal IC magnetic order, and
a sharp peak at the FE transition temperature, $T_{C1}$, where the magnetic order locks into a CM modulation. The sharp increase of
$\varepsilon(T)$ at $T_{C2}\simeq17 K$ indicates the unlocking of the magnetic modulation into the low-$T$ IC phase. The FE polarization
displays the corresponding increase at $T_{C1}$ and drop at $T_{C2}$, respectively (Fig. 1). An additional small step-like anomaly of
$\varepsilon(T)$ at $T_{N2}\simeq19 K$ that is also reflected in a sudden change of slope of the FE polarization upon cooling has not been
reported before. It is similar to the more pronounced step of $\varepsilon(T)$ observed in $HoMn_2O_5$.\cite{delacruz:06} Note that our data for
$P(T)$ confirm the sign change of the FE polarization at $T_{C2}$ although the value of $P(T)$ in the CM phase ($T_{C2}<T<T_{C1}$) is
significantly different from earlier reports. This is due to the special procedure of poling the FE domains used in the present work which
provides a more accurate determination of $P(T)$.\cite{delacruz:08}

With the application of hydrostatic pressure the FE polarization $P$ is dramatically altered at low temperatures (Fig. 2a). At low pressure
$P(T)$ increases slightly but it suddenly changes sign at a critical pressure of $p_c\simeq$10 kbar. After the sign reversal $P(T)$ increases
quickly and approaches the FE polarization of the CM high-polarization phase. At $T_{C2}$ a small drop of $P(T)$ is still visible up to 14 kbar,
but it disappears completely at the highest pressure of this investigation, 16.8 kbar suggesting that the IC low-temperature phase is completely
suppressed by pressures exceeding 14 kbar. The unique pressure-induced sign reversal of $P$ has not been observed before and it needs a careful
discussion of the magnetic structures and its relations to the ferroelectric order. The temperature dependence of the dielectric constant at
various pressures (Fig. 2b) supports the conclusions derived from the polarization measurements. The sharp increase of $\varepsilon(T)$ at
$T_{C2}$ is enhanced with the application of pressure up to about 10 kbar (indicated by the solid arrow in Fig. 2b). However, above the critical
pressure $\varepsilon(T)$ of the low-$T$ IC phase suddenly decreases as the polarization changes sign (dotted arrow in Fig. 2b). The pressure
related changes of $\varepsilon$ and $P$ in the low-$T$ IC phase are clearly visible in Fig. 3 showing both, $\varepsilon(5 K)$ and $P(5 K)$ as
functions of pressure. The transition from negative to positive FE polarization happens in a narrow pressure range between 9.5 and 10.5 kbar. In
the same pressure range $\varepsilon(5 K)$ experiences a sharp drop. It is interesting that the phase transition at $T_{C2}$ is only suppressed
at a much higher pressure suggesting that the sign change of the FE polarization (at 10 kbar) and the ICM $\rightarrow$ CM transition (above 14
kbar) are two well distinguished phenomena that have to be considered as separate transitions. The maximum of the low-$T$ $\varepsilon(p)$ in
Fig. 3 indicates a softness of the lattice in response to the electric field right at the critical pressure where the magnetic structure shows a
significant change. This is further evidence for the strong coupling of the spins to the lattice. When the magnetic system becomes less rigid at
the transition the lattice follows and the response to an electric field (dielectric constant) exhibits a maximum.

To understand the effect of external pressure on the ferroelectricity in $YMn_2O_5$ better we investigated the subtle changes of the lattice at
the various phase changes. While x-ray scattering experiments have not been successful in resolving the lattice strain at the FE
transitions\cite{kagomiya:03} our high-resolution thermal expansion measurements clearly show the abrupt change of the lattice constants at all
magnetic and FE transitions (Fig. 4). Upon decreasing $T$ the $a$- and $c$-axes show a sudden increase at $T_{N1}$ whereas there is no anomaly
in the $b$-axis length. This is associated with the onset of the IC magnetic modulation along $a$ and $c$. At the FE transition ($T_{C1}$) all
three axes indicate a small increase as a signature of the onset of FE order and the locking into the CM magnetic phase. The largest expansion
anomaly, however, is detected at $T_{C2}$, the low-$T$ ICM $\rightarrow$ CM transition where the FE polarization reverses sign. The relative
change of $a$, $b$, and $c$ at $T_{C2}$ are $9.3*10^{-6}$, $7.1*10^{-6}$, and $-19.2*10^{-6}$, respectively. While the volume change is small,
the strain on the lattice parameters is significant, $a$ and $b$ expand while $c$ shrinks in passing into the low-$T$ IC phase. The similar
change of the lattice constants at $T_{C2}$ in other $RMn_2O_5$ crystals ($R=Ho,Tb$)\cite{delacruz:06} imply the common origin of this typical
lattice distortion which is the partial release of the magnetic frustration between the $a$-axis zigzag chains.

The low-$T$ CM $\rightarrow$ ICM transition is supposedly driven by the increasing magnetic frustration between two adjacent $Mn$-spin chains
along the $a$-axis. The magnetic coupling between these two chains is mainly through superexchange interactions between pairs of $Mn^{3+}$ and
$Mn^{4+}$ spins via the neighboring oxygen ions. In the CM ferroelectric phase, following the zigzag chains along $a$, every second pair of
spins is frustrated whereas the remaining pairs are not.\cite{blake:05,chapon:06} With the decrease of temperature the magnitude of this
frustration increases with the increase of the $Mn$ sublattice magnetization and results in the phase transition at $T_{C2}$. There is a
combination of magnetic and lattice effects involved: (i) The lattice distortion as verified by the expansion anomalies (Fig. 4) does modify the
relevant exchange coupling constants. In particular, the in-plane expansion and the $c$-axis compression slightly increase the angle of the
$Mn^{3+}-O-Mn^{4+}$ superexchange coupling decreasing the magnitude of the corresponding coupling constant (J3 in the notation of Blake et
al.\cite{blake:05}) and the associated spin frustration. The Mn ions of the frustrated pair are also displaced to further reduce their exchange
energy.\cite{kagomiya:03} (ii) According to neutron scattering data\cite{blake:05,chapon:06} the local spin orientation also changes
dramatically in the low-$T$ IC phase increasing the relative angle between the spins of two neighboring chains to about
40$^\circ$.\cite{chapon:06} The large angle between the spin systems reduces the frustration further but it also decreases the overall coupling
between the spin chains. (iii) As a consequence, the magnetic modulation of every second chain "slides" by increasing the relative phase with
respect to the magnetic wave of adjacent chains.\cite{chapon:06} This change of phase results in a further decrease of both, the magnetic
frustration and the FE polarization, as e.g. in $HoMn_2O_5$ and $TbMn_2O_5$, or even in a reversal of the polarization (with a decrease in
magnitude) as in $YMn_2O_5$. Simultaneously, the magnetic modulation along $a$ and $c$ unlock from their commensurate values and the low-$T$
phase becomes incommensurate.

The application of pressure, as shown in Fig. 2, reverses the polarization and stabilizes the commensurate phase. This is obviously triggered by
the compression of the $a$-$b$ plane suggesting that the FE polarization could also be controlled by an in-plane strain induced in thin films of
$YMn_2O_5$ on appropriate substrates at ambient pressures. At the first critical pressure of 10 kbar, the phase shift of the magnetic order in
neighboring chains is reduced which results in the reversal of the polarization from $P<0$ to $P>0$.\cite{chapon:06} However, the sudden drop of
the $P(T)$ at $T_{C2}$ still persists above 10 kbar and the polarization smoothly increases with raising pressure. Only above the second
critical pressure of 14 kbar the IC phase becomes unstable and transforms into the high-polarization commensurate phase.

These results provide convincing evidence for the dramatic effects of lattice strain and the strength of the spin-lattice coupling on the
ferroelectricity in $YMn_2O_5$. The data for $YMn_2O_5$ should be compared with similar data for $HoMn_2O_5$.\cite{delacruz:06,delacruz:06b}
This is an excellent opportunity to reveal the potential role of the rare earth moment in these compounds. While the ionic size of the $Y^{3+}$
and $Ho^{3+}$ ions is nearly identical the major difference between both compounds is the existence of the $Ho$ magnetic moment that is missing
in $YMn_2O_5$. The phase sequence of both compounds is surprisingly similar. Even the spin re-orientation transition at $T_{N2}$ in the
commensurate phase previously identified in $HoMn_2O_5$\cite{delacruz:06} is evident in $YMn_2O_5$ in the form of the small step of
$\varepsilon(T)$ and the increase of $P(T)$ at 20 K with decreasing $T$. The main difference between both compounds is the magnitude of the
polarization drop at $T_{C2}$. In $HoMn_2O_5$ the polarization decreases but remains positive whereas in $YMn_2O_5$ the drop is so large that
$P$ reverses sign in the low-$T$ IC phase. Within the exchange striction model the drop of $P(T)$ can be explained by the change of spin
orientation in every second chain and by the phase shift between the magnetic modulation of neighboring chains. It appears conceivable that the
presence of the $Ho$ magnetic moments and their interaction with the $Mn$ moments stiffens the $Mn$ chain system and reduces the phase shift
between adjacent chains. This explains the lesser degree of the polarization change at $T_{C2}$ (still remaining positive) in $HoMn_2O_5$ as
compared to $YMn_2O_5$ (sign reversal). Detailed neutron scattering experiments on $HoMn_2O_5$, similar to the investigation of $YMn_2O_5$, are
suggested to reveal the details of the magnetic orders in passing from the commensurate into the incommensurate phase at $T_{C2}$ and to confirm
our conclusions.

\begin{acknowledgments}
This work is supported in part by the T.L.L. Temple Foundation, the J. J. and R. Moores Endowment, and the State of Texas through TCSUH. Work at
Rutgers was supported by NSF-DMR-0520471.
\end{acknowledgments}

\bibliographystyle{phpf}


\begin{figure}
\caption{(Color online) Dielectric constant (circles: decreasing T, squares: increasing T) and ferroelectric polarization (dashed lines) of
$YMn_2O_5$. $T_{N2}$ and $T_{C2}$ are marked for decreasing $T$ only.} \caption{(Color online) Temperature dependence of (a) the ferroelectric
polarization and (b) the dielectric constant of $YMn_2O_5$ at different pressures.} \caption{(Color online) Dielectric constant and
ferroelectric polarization of $YMn_2O_5$ at 5 K as function of pressure.} \caption{(Color online) Temperature dependence of lattice constants of
$YMn_2O_5$ below 50 K.}
\end{figure}

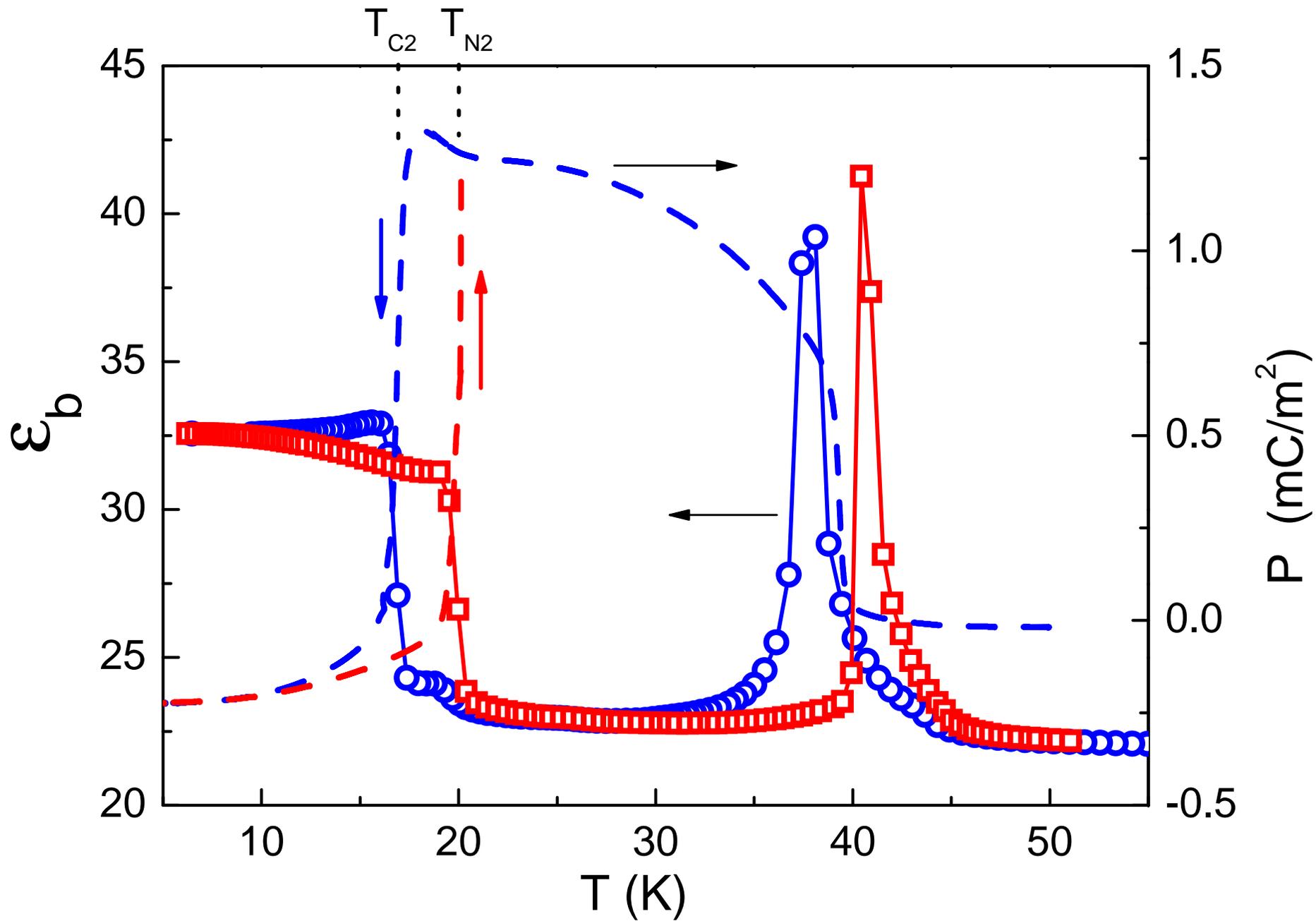
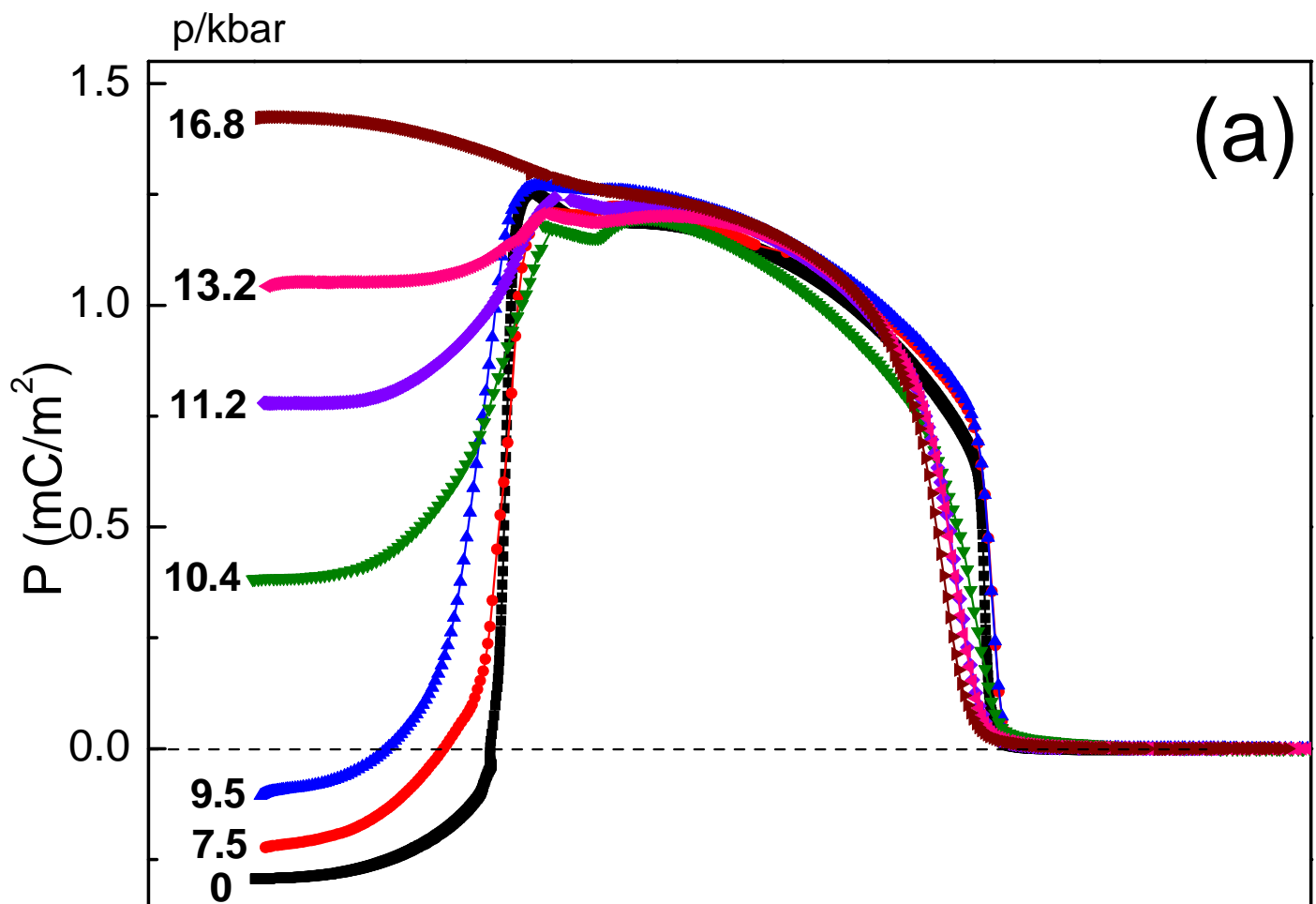
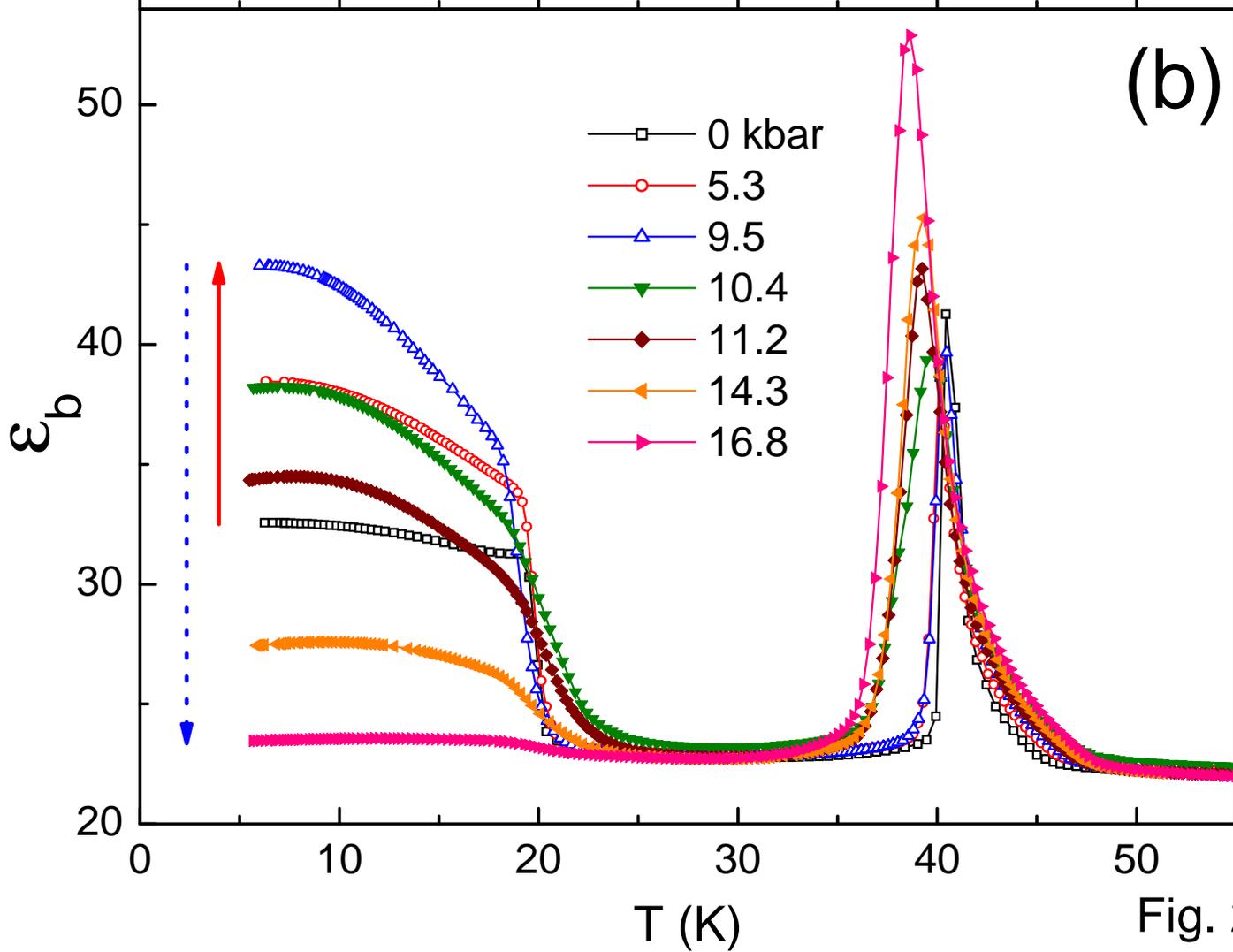
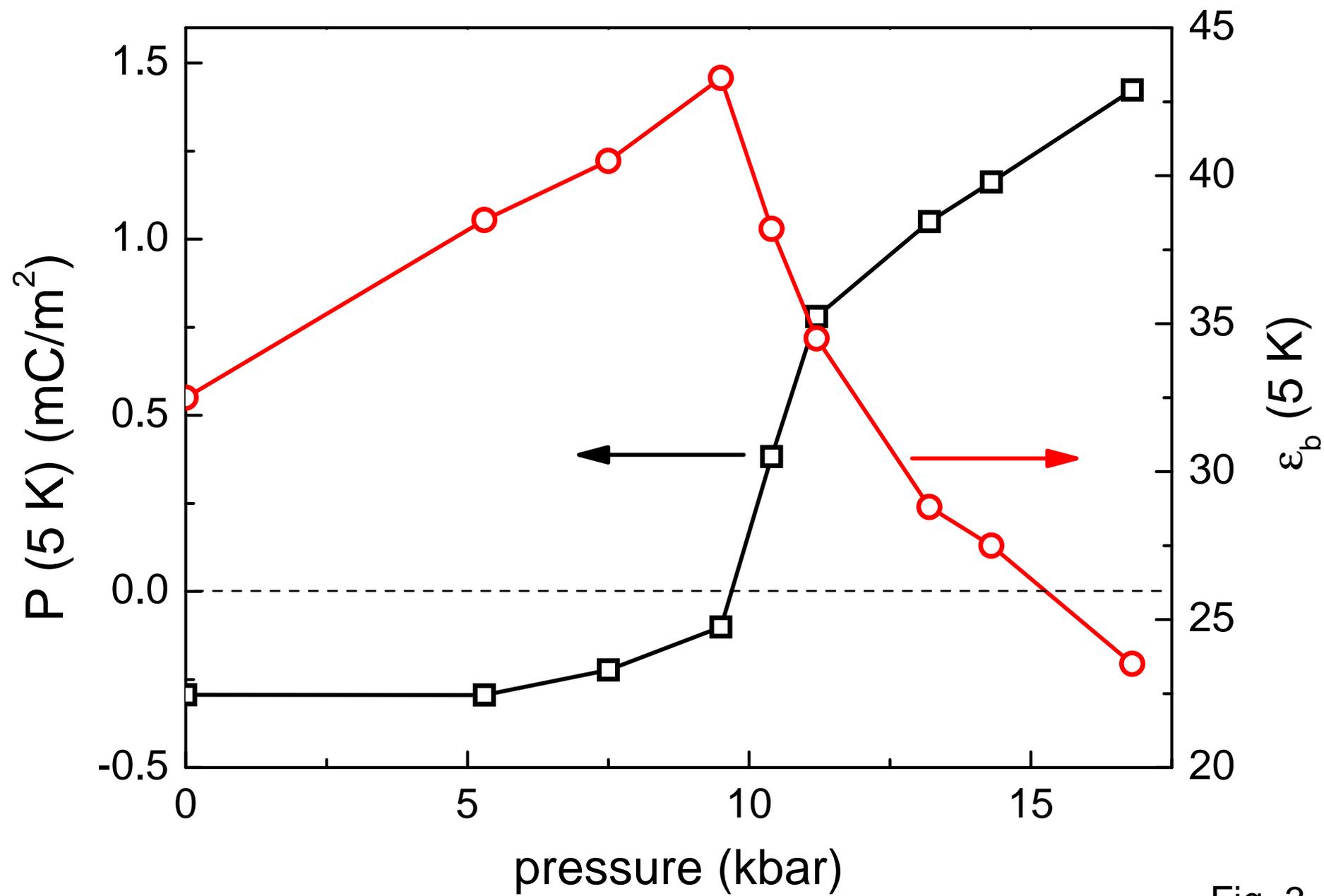
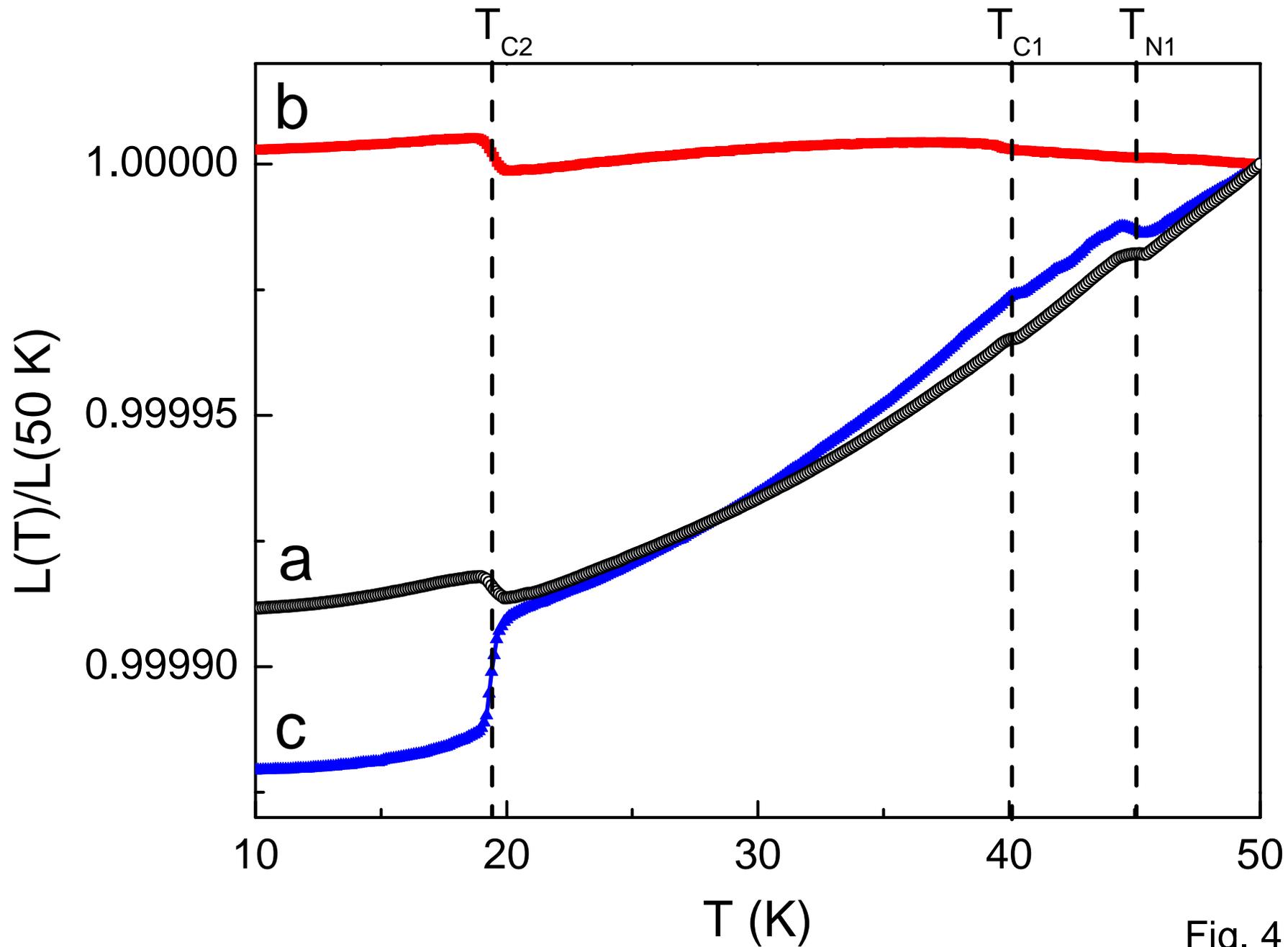
\end{document}